\documentclass[twocolumn]{jpsj3}
\usepackage{txfonts}

\usepackage{color}
\usepackage{ulem}

\usepackage{bm}

\title{Magnetic Field--Temperature Phase Diagram of CeCoSi Constructed on the Basis of Specific Heat, Magnetoresistivity, and Magnetization Measurements: Single Crystal Study}

\author{Hiroyuki Hidaka$^1$\thanks{hidaka@phys.sci.hokudai.ac.jp}, Shun Yanagiya$^1$, Eikai Hayasaka$^1$, Yuma Kaneko$^1$, Tatsuya Yanagisawa$^1$, \mbox{Hiroshi Tanida$^2$}, and Hiroshi Amitsuka$^1$}

\inst{$^1$Graduate School of Science, Hokkaido University, Sapporo, Hokkaido 060-0810, Japan \\
$^2$Liberal Arts and Sciences, Toyama Prefectural University, Toyama 939-0398, Japan
} 

\abst{The Ce-based metallic compound CeCoSi with a tetragonal structure exhibits successive phase transitions: a hidden ordering, whose order parameter is unidentified, at $T_{\rm 0}$ $\sim$ 12 K and an antiferromagnetic ordering at $T_{\rm N}$ = 9.4 K. [H. Tanida $et$ $al$., J. Phys. Soc. Jpn. ${\bf 88}$, 054716 (2019).] 
We performed specific heat, magnetoresistivity (MR), and magnetization ($M$) measurements for single crystals of CeCoSi at low temperatures in magnetic fields $B$ of up to 14 T and constructed detailed magnetic field--temperature phase diagrams for both $B$ $\parallel$ [100] and [001]. 
In the longitudinal MR measurements for $B$ $\parallel$ [100], the sign of MR changes from negative in a paramagnetic state to positive in the hidden ordered state, which suggests a change in electronic state across $T_{\rm 0}$.  
In addition, the longitudinal MR and magnetization processes for $B$ $\parallel$ [100] show anomalies upon the application of $B$ in both ordered states, suggesting the presence of $B$-induced regions. 
Similar $B$-induced regions may exist in the magnetic phase diagram for $B$ $\parallel$ [001]. 
The emergence of $B$-induced regions may be attributed to a change in the symmetry of the order parameter or domain alignment on applying $B$. 
}

\begin{document}
\maketitle

\section{Introduction}

An ordered state in solids has been clarified by revealing changes in the symmetries of the relevant system, such as point-group, translational, and time-reversal symmetries, accompanying the phase transition. 
However, there is an enigmatic ordering whose order parameter (OP) is unidentified, called a ``hidden order (HO)", in several materials, such as URu$_2$Si$_2$ \cite{URu2Si2-1,URu2Si2-2}, and studies to reveal its OP are being intensively carried out.  
In recent years, the presence of an ordering of multipole degrees of freedom has been established both experimentally and theoretically through attempts to elucidate the HO. 
Here, the multipole is characterized by the anisotropic distribution of electric or magnetic charge. \cite{Hayami, Watanabe}  
The 4$f$ electron system \mbox{CeCoSi} has also recently been reported to show the HO. \cite{Tanida_poly, Tanida_single}
The HO in CeCoSi has attracted considerable attention because of lack of local space-inversion symmetry at Ce sites \cite{Bodak}, where an odd-parity multipole, in addition to an even-parity one, is expected to play a particular role. \cite{Yatsushiro}

CeCoSi crystallizes in the tetragonal CeFeSi-type structure with the space group $P$4/$nmm$ (No. 129, $D_{4h}^{7}$) \cite{Bodak}, where the space-inversion symmetry at  Ce sites is locally broken. 
This compound exhibits two successive phase transitions: an antiferromagnetic (AFM) ordering at $T_{\rm N}$ = 9.4 K and the HO at $T_{\rm 0}$ $\sim$ 12 K. \cite{Tanida_poly, Tanida_single, Chevalier} 
Recent neutron diffraction measurements suggested that Ce moments with a size of 0.37(6) $\mu_{\rm B}$/Ce are aligned along the [100] direction with a \mbox{\boldmath $q$} = 0 structure in the AFM state \cite{Nikitin}. 
On the other hand, the possibility of an antiferroquadrupole (AFQ) ordering of Ce 4$f$ electrons has been suggested as a candidate for the HO below $T_{\rm 0}$ from the fact that $T_{\rm 0}$ increases with the magnetic field $B$, at least up to 7 T. \cite{Tanida_poly, Tanida_single} 
The Co nuclear quadrupole resonance (NQR) measurements for a single-crystalline \mbox{CeCoSi} revealed that the 4-fold and time-reversal symmetries are preserved at Co sites below $T_{\rm 0}$, suggesting the AFQ or higher electric multipole ordering. \cite{Manago} 
However, a Kramers doublet crystalline-electric-field (CEF) ground state (GS) without the quadrupole degrees of freedom and CEF excited states located at approximately 100 K seem to be incompatible with the AFQ ordering at $T_{\rm 0}$ of  \mbox{12 K}. \cite{Tanida_single, Nikitin, Comment_Yatsushiro} 
Furthermore, an obscure anomaly in bulk properties at $T_{\rm 0}$ \cite{Tanida_single} is different from that observed in conventional AFQ ordering compounds. \cite{CeB6_C, CeB6_rho, PrPb3_Tayama, TmTe} 
In addition, very recent X-ray diffraction (XRD) measurements on a single-crystalline CeCoSi revealed that a triclinic lattice distortion occurs below $T_{\rm 0}$ \cite{XRD_single}, and its relationship with the results of the NQR measurements must be clarified. 
Thus, it is essential to obtain a detailed magnetic field--temperature ($B$--$T$) phase diagram of CeCoSi up to high magnetic fields, including its anisotropy with respect to the $B$ direction, because the magnetic phase diagram will provide fundamental and important information about the ordered states.

In the present study, we have performed specific heat $C$, electrical resistivity $\rho$, and magnetization $M$ measurements for single crystals of CeCoSi in high magnetic fields up to 14 T and constructed detailed $B$--$T$ phase diagrams for $B$ applied parallel to the tetragonal [100] and [001] axes. 
The experimental results are presented in Sect. 3; the $B$--$T$ phase diagram in Sect. 3.1, $C$ in Sect. 3.2, $\rho$ in Sect. 3.3, and $M$ in Sect. 3.4. 
In addition, we discuss the characteristics of the obtained magnetic phase diagrams in Sect. 4.

\section{Experimental Details}

Single crystals of CeCoSi were prepared by a Ce/Co eutectic flux method, as described in the previous paper \cite{Tanida_single}. 
A sample piece, named $\#$1, was used for all the measurements, while another sample piece taken from a different batch, named $\#$2, was used only for the $M$ measurements to examine the reproducibility. 
The dimensions of the thin plate-like samples $\#$1 and $\#$2 are $\sim$ 2.0 $\times$ 1.0 $\times$ 0.2 and $\sim$ 2.0 $\times$ 2.0 $\times$ 0.4 mm$^3$ for the tetragonal [100]--[010]--[001] axes, respectively.  

The $C$ measurements were performed by a thermal-relaxation method in the temperature range of 2--300 K at magnetic fields of $B$ = 0 and 6 T using a physical property measurement system (PPMS; Quantum Design). 
The magnetic field $B$ was applied along [100]. 
The $\rho$ measurements were performed by a conventional four-probe method in the temperature range of 2--300 K and in the magnetic field range of --14 -- 14 T using the PPMS. 
We measured the magnetoresistivity (MR) in two different configurations: a longitudinal MR with electric current \mbox{$J$ $\parallel$ [100]} and $B$ $\parallel$ [100], and a transverse MR with $J$ $\parallel$ [100] and $B$ $\parallel$ [001]. 
The DC $M$ measurements were performed in the temperature range of 2--300 K and in the magnetic-field range of 0--7 T using a magnetic property measurement system (MPMS; Quantum Design). 
Magnetic fields were applied along the [100] and [001] directions. 
The obtained $M$ data for sample $\#$2 are shown in the Sect. 4 of Supplemental Materials (SM). \cite{SM}

\section{Experimental Results}
\subsection{Magnetic field--temperature phase diagram}

\begin{figure}[tb]
\begin{centering}
\includegraphics[width=0.4\textwidth]{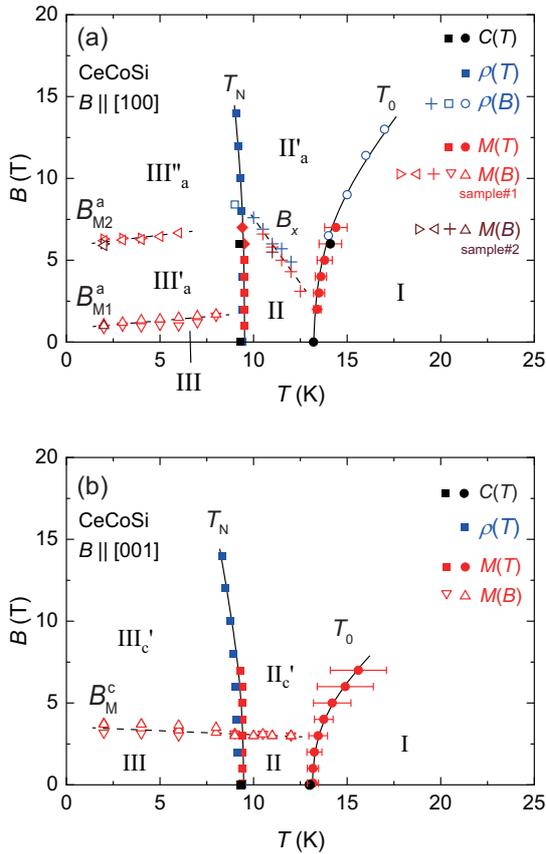}
\caption{
(Color online) Magnetic field--temperature phase diagrams of \mbox{CeCoSi} constructed on the basis of the present measurements of sample $\#$1 for (a) $B$ $\parallel$ [100] and (b) $B$ $\parallel$ [001].  
In (a), the data obtained from the $M$ measurements of sample $\#$2 are also plotted. 
The upward and rightward triangles were determined during the $B$ increasing process in $M$($B$), while the downward and leftward triangles were determined during the $B$ decreasing process. 
The solid and dashed curves are guides for the eye. 
}
\label{Fig1} 
\end{centering}
\end{figure}

First, we show the $B-T$ phase diagrams of CeCoSi constructed on the basis of the present $C$, $\rho$, and $M$ measurements for \mbox{$B$ $\parallel$ [100]} and [001] in Figs. 1(a) and 1(b), respectively.  
The details of each measurement will be given later. 
The main feature is that these phase diagrams have $B$-induced regions in the ordered states for $B$ $\parallel$ [100] and possibly also for [001].  
Here, we refer to the regions of the HO phase as II and ${\rm II'}_a$ for $B$ $\parallel$ [100], and as II and ${\rm II'}_c$ for $B$ $\parallel$ [001], in order from the lowest magnetic field.  
In addition, we refer to the regions of the AFM phase as III, ${\rm III'}_a$, and ${\rm III''}_a$ for $B$ $\parallel$ [100], and as III and ${\rm III'}_c$ for $B$ $\parallel$ [001]. 
Region I corresponds to a paramagnetic (PM) phase.

\subsection{Specific heat}
Figure 2 shows the temperature dependence of $C_{4f}$ in CeCoSi (sample $\#$1) at $B$ = 0 and 6 T applied parallel to [100]. 
Here, $C_{4f}$ is the contribution of 4$f$ electrons to the specific heat, which was estimated by subtracting $C$ of LaCoSi at zero field from $C$ of CeCoSi. \cite{Tanida_single} 
The obtained features of $C_{4f}$($T$), which are a $\lambda$-type anomaly at $T_{\rm N}$, a small jump at $T_{\rm 0}$, and a CEF Schottky anomaly at around 40 K, are in good agreement with those reported previously. \cite{Tanida_single} 
$T_{\rm 0}$ at zero field is estimated to be $\sim$13 K in the present sample, which is slightly higher than that of 12 K reported previously. \cite{Tanida_single, Comment_Nikitin}
The blue curve represents the calculated CEF Schottky curve based on the previously suggested CEF level scheme. \cite{Nikitin} 
Applying $B$ of 6 T negligibly affects $C_{4f}$($T$) except for a slight decrease in $T_{\rm N}$ and an increase in $T_{\rm 0}$.   
The magnitude of the jump at $T_{\rm 0}$ also changes negligibly despite the increase in transition temperature from 13 to 14 K, as shown in the inset of Fig. 2. 
The obtained $T_{\rm 0}$ and $T_{\rm N}$ are plotted in the $B$--$T$ phase diagram [Fig. 1(a)]. 
Note that the present $C$ measurement at \mbox{6 T} does not show any clear anomaly at $B_x$, which separates the II and ${\rm II'}_a$ regions.

\begin{figure}[tb]
\begin{centering}
\includegraphics[width=0.4\textwidth]{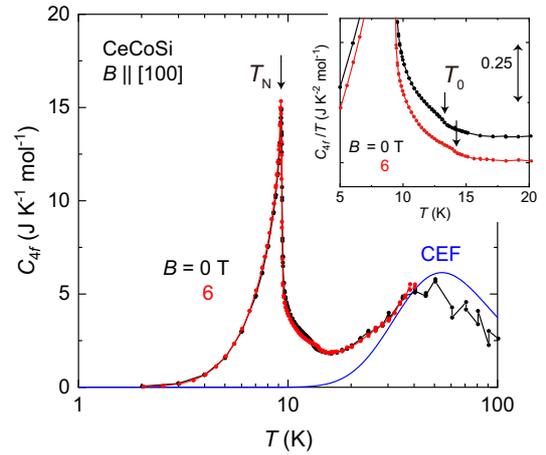}
\caption{
(Color online) Temperature dependence of $C_{4f}$ in CeCoSi at 0 and 6 T for $B$ $\parallel$ [100]. 
The blue curve represents the calculated CEF Schottky peak. \cite{Nikitin}
The inset shows the enlarged view of $C_{4f}$($T$)/$T$ near $T_{\rm 0}$, where the  data are shifted vertically for ease of viewing. 
}
\label{Fig2} 
\end{centering}
\end{figure}

\subsection{Electrical resistivity} 

Figures 3(a) and 3(b) show the temperature dependence of the electrical resistivity $\rho$($T$) below 30 K at magnetic fields up to 14 T for CeCoSi (sample $\#$1) with electric current $J$ applied along the [100] direction and magnetic fields applied along the [100] and [001] directions, respectively. 
The $\rho$($T$) data at zero field shows no clear anomaly at $T_{\rm 0}$, although $\rho$ seems to decrease slightly below $T_{\rm 0}$, which is determined from the present $C$($T$) measurement. 
On the other hand, $\rho$($T$) shows a clear kink anomaly at $T_{\rm N}$. 
The effect of $B$ on $T_{\rm 0}$ is obscure in both the $\rho$($T$) measurements, while the decrease in $T_{\rm N}$ for $B$ $\parallel$ [001] is greater than that for $B$ $\parallel$ [100]. 
These results at zero field are consistent with the previous report. \cite{Tanida_single} 
In addition, the small residual resistivity of \mbox{$\sim$3 $\mu$$\Omega$cm} guarantees the high quality of the present sample.

\begin{figure}[tb]
\begin{centering}
\includegraphics[width=0.35\textwidth]{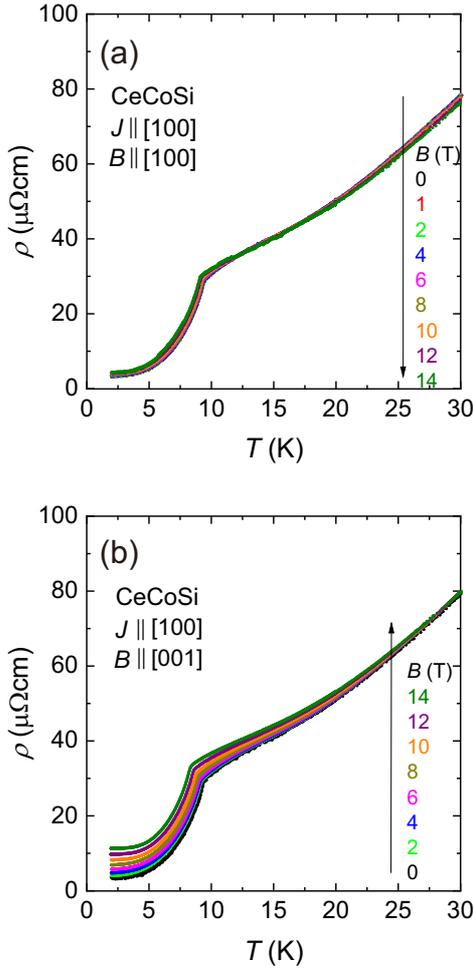}
\caption{
(Color online) 
Temperature dependence of electrical resistivity $\rho$($T$) in CeCoSi (sample $\#$1) in magnetic fields up to 14 T for (a) \mbox{$B$ $\parallel$ [100]} and (b) \mbox{$B$ $\parallel$ [001]}.
Electric current $J$ was applied along the [100] axis in both measurements. 
}
\label{Fig3}
\end{centering}
\end{figure}

When the magnetic field is applied along the [100] direction, the $\rho$($T$) curve seems to be nominally affected by $B$, as can be seen in Fig. 3(a).  
However, in the longitudinal MR measurements for $J$ and $B$ $\parallel$ [100], the $B$ dependence of $\rho$ shows a clear kink when crossing the phase boundary between the PM and HO phases. 
Figure 4(a) shows $\Delta\rho$($B$)/$\rho$(0) obtained from the longitudinal MR measurements in CeCoSi for $J$ and \mbox{$B$ $\parallel$ [100]} at several temperatures between 10 and 20 K in the field range up to +14 T. 
The data are shifted vertically for clarity in this figure. 
Here, $\Delta\rho$($B$) and $\rho$(0) are defined as \mbox{[$\rho$($B$) -- $\rho$(0)]} and $\rho$ at $B$ = 0 at a fixed temperature, respectively. 
At 20 K, which corresponds to the PM state, $\Delta\rho$($B$)/$\rho$(0) shows a monotonic decrease with increasing $B$.  
Below 17 K, one can see a kink anomaly at a high field of $\sim$13 T. 
The corresponding field of the kink anomaly is defined as $B_{\rm 0}$ [blue open circle in Fig. 1(a)], since $B_{\rm 0}$ decreases with decreasing temperature down to 14 K and then appears to merge with the $T_{\rm 0}$($B$) curve in the $B$--$T$ phase diagram.

It is characteristic that the sign of MR in the low $B$ region appears to change gradually from negative in the PM state to positive in the HO state below $T_{\rm 0}$ of $\sim$13 K with decreasing temperature. 
The negative MR in the PM state should be due to the suppression of magnetic scattering between the 4$f$ and conduction electrons by the external $B$. 
On the other hand, the positive MR observed below 13 K would be derived from an electronic state in the HO state. 
The electronic state in the HO state should be different from that in the PM state, since the XRD measurements have revealed the HO of CeCoSi to be accompanied with a structural change. \cite{XRD_single} 
An anomaly in $\rho$($T$) at $T_{\rm 0}$ under high pressure, which is suggestive of gap opening, also suggests a change in electronic state because of the HO. \cite{Tanida_poly , Lengyel} 
If the positive MR in the HO state mixes with the negative MR in the PM state from above $T_{\rm 0}$, the gradual sign change observed in the present study can be explained. 
This explanation is supported by the XRD measurements, where an increase in structural fluctuation is suggested with decreasing temperature from $\sim$20 K toward $T_{\rm 0}$. \cite{XRD_single}  
In addition, we found another anomaly above $\sim$5 T below $T_{\rm 0}$, which increases with decreasing temperature, as can be seen in Fig. 4(a). 
This result suggests the presence of a high-$B$ region, named ${\rm II'}_a$, above region II. 
Here, the corresponding field of the newly found anomaly below $T_{\rm 0}$ is defined as $B_x$.

The results of $\Delta\rho$($B$)/$\rho$(0) for $B$ $\parallel$ [100] in the AFM state are shown in Fig. 4(b).  
These $\Delta\rho$($B$)/$\rho$(0) curves show the positive MR, although the curve changes gradually from downward to upward convex with decreasing temperature. 
The positive MR in the AFM state may also explain the sign change in MR observed at around $T_{\rm 0}$, since the presence of weak AFM fluctuations has been suggested above $T_{\rm N}$ by the NQR measurements \cite{Manago}.   
However, the temperature at which the AFM fluctuations develop ($\sim$12 K) seems low to account for the present sign change in MR. 
The anomaly observed at \mbox{9 K} corresponds to the transition field from the AFM phase to the HO phase, defined as $B_{\rm N}$ [blue open square in Fig. 1(a)]. 
None of the clear anomalies corresponding to $B_{M1}^a$ and $B_{M2}^a$, which were observed in the $M$($B$) measurements shown in Sect. 3.4, were observed in the present MR measurements.

\begin{figure}[tb]
\begin{centering}
\includegraphics[width=0.45\textwidth]{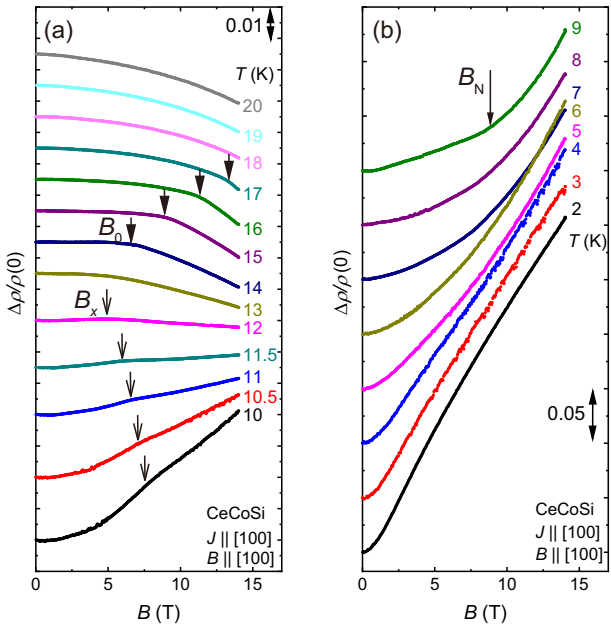}
\caption{(Color online) $\Delta\rho$/$\rho(0)$ in longitudinal MR for CeCoSi (sample $\#$1) plotted as a function of $B$ at several temperatures in the range (a) between 10 and 20 K and (b) below 9 K. 
Both $J$ and $B$ were applied along the [100] direction. 
The data are shifted vertically for ease of viewing. 
}
\label{Fig4}
\end{centering}
\end{figure}

\begin{figure}[tb]
\begin{centering}
\includegraphics[width=0.4\textwidth]{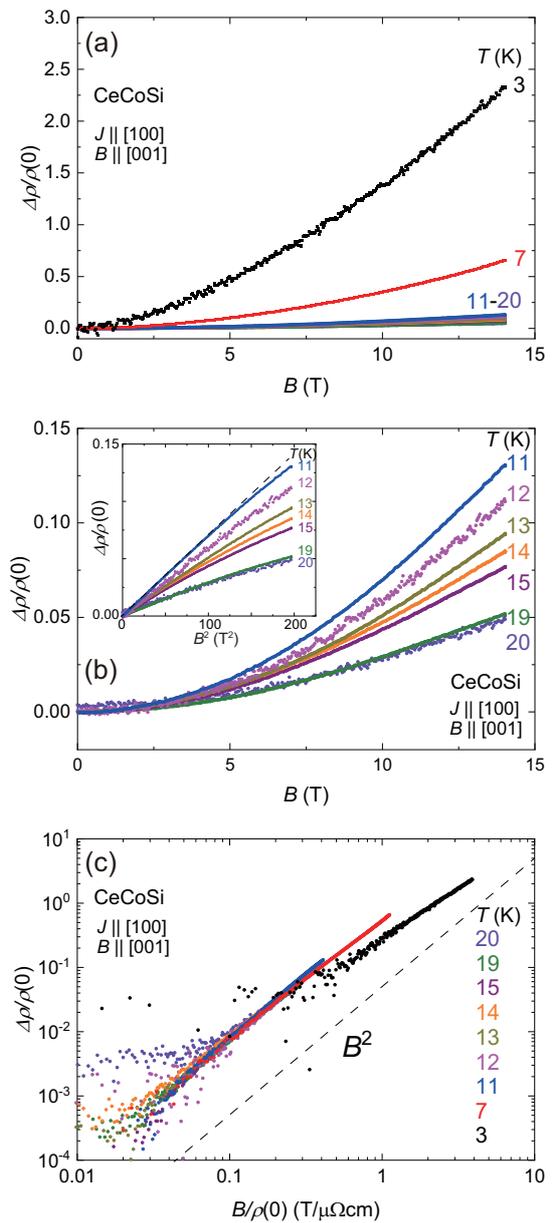}
\caption{(Color online) 
$\Delta\rho$/$\rho(0)$ in the transverse MR for CeCoSi (sample $\#$1) plotted as a function of $B$ at several temperatures (a) below  20 K and (b) between 11 and 20 K. 
Here, $J$ was applied along the [100] direction, and $B$ was applied along the [001] direction. 
The inset of (b) shows the $B^2$ dependence of $\Delta\rho$/$\rho(0)$ between 11 and 20 K. 
(c) shows their Kohler plots at all the measured temperatures. 
The dashed lines in the inset of (b) and in (c) represent $B^2$ dependence. 
}
\label{Fig5}
\end{centering}
\end{figure}

Figures 5(a) and 5(b) show the $\Delta\rho$($B$)/$\rho$(0) curves obtained from the transverse MR measurements for $J$ $\parallel$ [100] and \mbox{$B$ $\parallel$ [001]} below 20 K and between 11 and 20 K, respectively. 
The raw data measured from $B$ = --14 to 14 T are slightly asymmetric with respect to $B$ because of a mixed Hall effect component that is approximately proportional to $B$. 
Here, we display the data after the Hall effect component was removed in Fig. 5.  
The raw $\rho$($B$) data and the estimated Hall effect are shown in Sect. 1 of SM. \cite{SM} 
The $\Delta\rho$($B$)/$\rho$(0) curves for $B$ $\parallel$ [001] exhibit parabolic behavior with the positive sign at all the investigated temperatures, and no kink anomaly or sign change is observed, which is in contrast to the results of $\Delta\rho$($B$)/$\rho$(0) curves in the longitudinal MR for $B$ $\parallel$ [100].

To see the characteristics of the transvers MR in CeCoSi, a Kohler plot, $\Delta\rho$($B$)/$\rho$(0) vs $B$/$\rho$(0), is shown in Fig. 5(c). 
For conventional metals, all Kohler plots from different temperatures are expected to fall on the same curve at low temperatures. \cite{WTe2, Pippard}  
The present results, except for the data below $T_{\rm N}$, satisfy the Kohler's rule and show  $B^2$ dependence in this \mbox{log--log} plot, although $\Delta\rho$($B$)/$\rho$(0) deviates from the $B^2$ behavior in the high $B$ region above $\sim$10 T, as shown in the inset of Fig. 5(b). 
Such $B^2$ dependence in the transverse MR can be explained by the scattering due to Lorentz force. \cite{Omoidensikei, Tanuma} 
In addition, the satisfaction of Kohler's rule appears to indicate that there is no marked change in the scattering mechanism of carriers even across $T_0$ within experimental accuracy; however, it cannot be concluded only from this feature that there is no change in the scattering mechanism between the PM and HO states.  
The slight change in the MR associated with the HO might be buried in the large positive MR due to Lorentz force, since the amount of change in the transverse MR for $B$ $\parallel$ [001] above $T_{\rm N}$ is several times larger than that in the longitudinal one for $B$ $\parallel$ [100], where the clear change is observed.  
The deviation from Kohler's rule and the $B^2$ behavior below $T_{\rm N}$ should be attributed to the change in scattering mechanism due to the AFM ordering.  
It is noted that a similar sign change in the MR has also been revealed by $\rho$ measurements under high pressure \cite{Lengyel} and for La-substitution systems \cite{Tanida_dope}, but it is difficult to compare those previous results with the present results, since those measurements were performed using polycrystalline samples.

\subsection{Magnetization} 

Figures 6(a) and 7(a) show the magnetization process $M$($B$) of CeCoSi (sample $\#$1) for $B$ $\parallel$ [100] and [001] up to 7 T, respectively, measured at temperatures between 2 and 30 K. 
The respective derivative $dM/dB$ values for $B$ $\parallel$ [100] and [001] are also shown in Figs. 6(b) and 7(b), respectively. 
The indicated data are shifted vertically for clarity in these figures. 
The steep decrease in $dM/dB$ near zero field, which is observed in all the derivative data, may be due to magnetic impurity. 
We also confirmed that the temperature dependence of the magnetic susceptibility [= $M$($T$)/$B$] for both the $B$ directions for sample $\#$1 are in good agreement with those reported previously, as shown in Sect. 2 of SM. \cite{SM, Tanida_single} 
The obtained transition temperatures $T_0$ and $T_{\rm N}$ are summarized in the $B$--$T$ phase diagrams (Fig. 1).

The $M$($B$) curves for $B$ $\parallel$ [100] seem to show simple linear behavior at all the measured temperatures; however, a slight deviation from the linear behavior can be seen at low temperatures, as seen in the derivative data [Fig. 6(b)]. 
The $dM/dB$ curves above 16 K are almost constant, whereas at 14 K it starts to increase at around $B_{\rm 0}$ determined from the present MR measurement. 
This behavior is consistent with the larger magnetic susceptibility in the HO state than in the PM state. 
Below $T_{\rm 0}$ $\sim$13 K, $dM/dB$ shows a broad maximum at the magnetic field corresponding to $B_x$. 
This result also suggests the presence of region ${\rm II'}_a$. 
All the $dM/dB$ curves measured between $T_{\rm N}$ and $T_{\rm 0}$ are shown in the Sect. 3 of SM. \cite{SM} 
In addition, the $dM/dB$ curves below $T_{\rm N}$ show two maximums at around 2 and 6 T accompanied by hysteresis. 
The hysteresis is observed only below $T_{\rm N}$. 
Here, the magnetic fields where these maximums appear are defined as $B_{M1}^a$ and $B_{M2}^a$, respectively [red open triangles in Fig. 1(a)]. 
$B_{M1}^a$ and $B_{M2}^a$ have different values between the $B$ increasing and decreasing processes because of the presence of the hysteresis. 
These features suggest the presence of $B$-induced regions, named ${\rm III'}_a$ and ${\rm III''}_a$, in the AFM ordered state.

A similar maximum in the $dM/dB$ curve is observed below $T_{\rm 0}$ even in the measurements with $B$ $\parallel$ [001], as shown in Fig. 7(b). 
In this measurement, the field where the maximum appears, named $B_M^c$, increases continuously as the temperature decreases. 
The single maximum above $T_{\rm N}$ changes to two maximums because of the hysteresis in the AFM state, similarly to the results for $B$ $\parallel$ [100]. 
Here, we named regions above $B_M^c$ in the phase diagram for $B$ $\parallel$ [001] regions ${\rm II'}_c$ and ${\rm III'}_c$, as can be seen Fig. 1(b).

To examine the reproducibility of the results of the $M$($B$) measurements using sample $\#$1, we performed additional $M$($B$) measurements using another sample piece taken from a different batch, named sample $\#$2. 
The measurement data of sample $\#$2 are shown in Sect. 4 of SM. \cite{SM} 
For \mbox{$B$ $\parallel$ [100]}, the anomalies at $B_x$, $B_{M1}^a$, and $B_{M2}^a$ are observed in the $dM/dB$ curves even for sample $\#$2, although the steep decrease in $dM/dB$ near zero field due to the magnetic impurity is more pronounced compared with the results for sample $\#$1. 
The hysteresis is also observed only in the AFM state. 
On the other hand, we cannot decide whether or not the $B_M^c$ anomaly is present from the measurements for $B$ $\parallel$ [001] using sample $\#$2 owing to the large effect of the magnetic impurity.  
The small hysteresis observed at 2 K might originate in the presence of $B_M^c$ anomaly. 
Therefore, we merely point out the possibility of the presence of $B$-induced regions in the phase diagram for $B$ $\parallel$ [001] at present.  
Measurements using samples with less impurity and other physical property measurements will be needed to verify the $B$-induced region.

\begin{figure}[tb]
\begin{centering}
\includegraphics[width=0.5\textwidth]{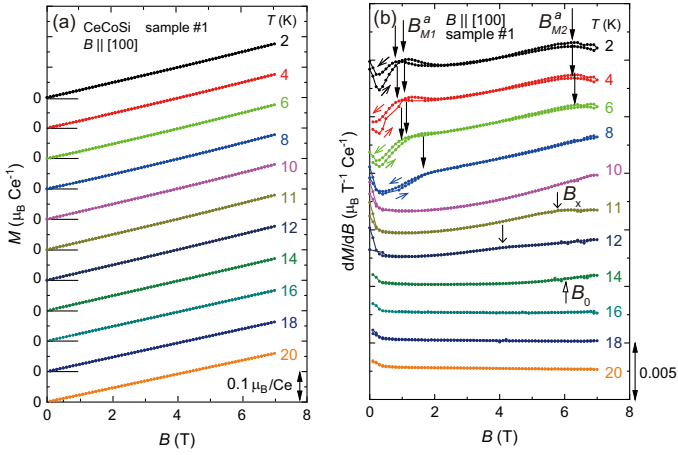}
\caption{(Color online) 
(a) Magnetization curves of CeCoSi (sample $\#$1) for \mbox{$B$ $\parallel$ [100]} at several temperatures below 20 K and (b) their respective derivatives $dM/dB$. 
The open arrow in (b) represents $B_{\rm 0}$ determined from the present MR measurement, and the colored arrows near the hysteresis represent the $B$ increasing and decreasing processes. 
The data are shifted vertically for ease of viewing. 
}
\label{Fig6}
\end{centering}
\end{figure}

\begin{figure}[tb]
\begin{centering}
\includegraphics[width=0.5\textwidth]{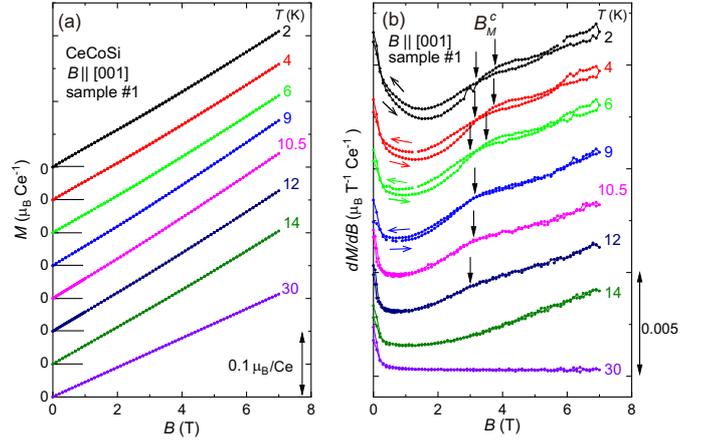}
\caption{(Color online) 
(a) Magnetization curves of CeCoSi (sample $\#$1) for \mbox{$B$ $\parallel$ [001]} at several temperatures below 30 K and (b) their respective derivatives $dM/dB$. 
The colored arrows near the hysteresis in (b) represent the $B$ increasing and decreasing processes.  
The data are shifted vertically for ease of viewing. 
}
\label{Fig7}
\end{centering}
\end{figure}

\section{Discussion}

We now discuss the characteristics of the $B$--$T$ phase diagrams constructed in the present study. 
First, $T_{\rm 0}$ increases and $T_{\rm N}$ decreases with increasing $B$ up to 14 T for both $B$ directions, where the $B$ effects on these ordering temperatures are larger for [001] than for [100]. 
Note that there is no direct experimental evidence to determine $T_{\rm 0}$ for [001] above 7 T. 
However, the HO phase for $B$ $\parallel$ [001] may be induced by $B$ lower than that for [100] at a fixed temperature above $T_{\rm 0}$, since $dM/dB$ for [001] at 14 K starts to increase from $B$ lower than that observed for [100] [see Figs. 6(b) and 7(b)].   
This anisotropic $B$ dependence of the ordering temperatures should be attributed to the symmetry of the OPs. 
Such an anisotropic increase in $T_{\rm 0}$ upon the application of $B$ is similar to a typical characteristic of the electric quadrupolar ordering regardless of whether it is a ferroquadrupolar (FQ) or an AFQ ordering. \cite{CeB6_rho, CeB6_C, PrPb3_Tayama, TmTe, Ce3-20-6, PrCu2} 
It has been determined that the critical field of the AFM state of CeCoSi is $\sim$20 T by high-field $M$ measurements for a polycrystalline sample. \cite{Tanida_single} 
On the other hand, it has not been clarified whether the $T_{\rm 0}$($B$) curve goes towards absolute zero or disappears under higher magnetic fields at present.  
This is an important issue to resolve for the elucidation of the ordered state of the HO phase in CeCoSi. 
Experiments under higher magnetic field are required.

Second, it is suggested that the HO phase of CeCoSi has the $B$-induced region for $B$ $\parallel$ [100] (and possibly [001]). 
Such a $B$-induced region would be associated with a change in the symmetry of OP or the alignment of domains by the application of $B$.  
Manago $et$ $al.$ pointed out the change in OP from hexadecapole or $O_{xy}$-type AFQ to $O_{zx}$-type AFQ in the results of the Co-NMR measurements at $B$ = 1 T applied along [100].  \cite{Manago}
Assuming a quadrupole ordering, it is considered that a stable state under high magnetic field is determined by the energy gain due to induced moments, such as magnetic dipole and magnetic octupole moments. \cite{Shiina} 
This possible change in the stable state would cause the suppression of the magnetic scattering in the MR and the small anomaly in the magnetization process observed in the present study. 
A similar anomaly in the MR accompanied by the emergence of the $B$-induced phase has also been reported in the AFQ ordering compound PrPb$_3$ \cite{PrPb3_Yoshida}, where the AFQ OP has been suggested to change from $O_2^0$ to $O_2^2$ by applying $B$ along the cubic [100]. \cite{PrPb3_Sato} 
Furthermore, it has been reported in the FQ compound PrTi$_2$Al$_{20}$ that the $dM$($B$)/$dB$ curve shows an anomaly originating from the discontinuous switching of the OP induced by the application of $B$. \cite{PrTi2Al20} 
It is notable that recent XRD measurements of single-crystalline CeCoSi have also revealed the presence of the $B$-induced region in the HO phase for $B$ $\parallel$ [100], where the triclinic lattice distortion is absent. \cite{XRD_single} 
Although it cannot be concluded from the present study that $B_x$ merges into $T_0$ at zero field, which was suggested by the results of the XRD study, the OP in region II may suppress its ordered state under magnetic field, while that in region ${\rm II'}_a$ may stabilize the state.

The $B$-induced regions are also observed in the AFM phase in CeCoSi. 
In the phase diagram for $B$ $\parallel$ [100], the AFM phase is separated into three regions by $B_{M1}^a$ and $B_{M2}^a$. 
The $B_{M1}^a$ anomaly should be independent of the presence of the ${\rm II'}_a$ region, while $B_{M2}^a$ appears to be connected to $B_x$ above $T_{\rm N}$.  
These $B$-induced anomalies in the AFM phase have also been pointed out from the previous XRD measurements, although the $B_{M2}^a$ determined by XRD ($\sim$8 T) is higher than that determined in the present study. \cite{XRD_single}
The XRD measurements suggested that preferable triclinic structural domains are selected at $B_{M1}^a$, and the triclinic distortion disappears above $B_{M2}^a$. \cite{XRD_single}
The hysteresis in the magnetization process, which is observed only in the AFM phase, suggests the presence of metastable states in the structural domain alignment. 
On the other hand, the possible presence of the ${\rm II'}_c$ and ${\rm III'}_c$ regions for $B$ $\parallel$ [001] might also be explained by the domain arrangement.  
Note that a similar magnetic phase diagram has been constructed on the basis of the XRD measurements for $B$ $\parallel$ [110]. \cite{XRD_single}

The magnetic structure of the AFM phase in CeCoSi is still controversial. 
The magnetic structure at zero field has been proposed to be a simple collinear one with the magnetic moments pointing along the [100] direction on the basis of the previous neutron powder diffraction measurements. \cite{Nikitin} 
However, considering the small anisotropy of the magnetic susceptibility \cite{Tanida_single}, the Co-NQR spectrum below $T_{\rm N}$, \cite{Manago} and the complex magnetic phase diagrams obtained in this study, the AFM magnetic structure may not be so simple one as proposed by Nikitin $et$ $al$. \cite{Nikitin}
Furthermore, if the antiferro-type ordering at Ce on-sites with \mbox{\boldmath $q$} = 0 occurs, the space-inversion symmetry at the midpoint between Ce ions is broken, resulting in the activation of the Dzyaloshinskii--Moriya (DM) interaction.  
The DM interaction will induce the tilting of magnetic moments in the AFM state. 
Taking these into consideration, the direction of magnetic moments is expected to be changed by the application of small $B$. 
The relationship between the possible change in the direction of magnetic moments and the alignment of the structural domains upon the application of $B$ also must be clarified.

Finally, the $B$--$T$ phase diagrams of CeCoSi could be explained by the successive phase transitions of the electric quadrupole and magnetic dipole orderings; however, the OP in the HO state has not yet been settled, including the antiferro or ferro type. 
Most compounds showing the quadrupole ordering of 4$f$ electrons have the quadrupole degrees of freedom in their CEF GS, such as a $\Gamma_8$ quartet for the Ce$^{3+}$ ion and a non-Kramers $\Gamma_3$ doublet for the Pr$^{3+}$ ion. \cite{CeB6_G8, PrPb3_Tayama, CeTe} 
On the other hand, since CeCoSi has the Kramers doublet CEF GS without the quadrupole degrees of freedom within the localized 4$f$-electron model ($J$ = 5/2), the quadrupole moment can be induced only between the GS and the excited state located at $\sim$100 K. \cite{Tanida_single, Nikitin} 
This CEF splitting seems to be too large to drive the quadrupole ordering of the local 4$f$ electron at $T_{\rm 0}$ of 13 K. 
The obscure anomaly at $T_{\rm 0}$ in the bulk properties of CeCoSi, such as $C$($T$), $M$($T$), and $\rho$($T$), is also a feature different from that found in the typical compounds showing the quadrupole ordering of the localized 4$f$ electron. \cite{CeB6_C, TmTe, PrPb3_Sato, PrPb3_Yoshida, CeAg_Morin} 
In particular, the jump at $T_{\rm 0}$ in $C$($T$)/$T$ of CeCoSi is considerably small, whereas the released entropy at $T_{\rm N}$ is large ($\sim$0.8 $R$ln2), suggesting the magnetic ordering of the 4$f$ electron. 
It is noteworthy that the magnitude of the jump at $T_{\rm 0}$ changes negligibly even at a  high magnetic field of 6 T. 
This robustness is in contrast to that in other AFQ compounds, where the jump in $C$($T$)/$T$ becomes larger as the ordered state stabilizes. \cite{CeB6_C, TmTe}

A higher-rank multipole than the quadrupole, which preserves the time-reversal symmetry \cite{Manago}, such as an electric hexadecapole, could also be a possible OP in the HO phase of CeCoSi, assuming the quasi-quartet CEF GS. 
Furthermore, an odd-parity multipole due to the lack of the local space-inversion symmetry at the Ce site might also play some role in this enigmatic ordering at $T_{\rm 0}$. \cite{Tanida_single, Yatsushiro} 
To clarify the ordered state in each region, including the high-field one, further studies, such as those on lattice distortion and the $B$--$T$ phase diagram under higher $B$ and high pressure, are needed and are now in progress.

\section{Summary}

We performed $C$, $\rho$, and $M$ measurements in single crystals of CeCoSi under $B$ up to 14 T applied parallel to the tetragonal [100] and [001] axes.  
For $B$ $\parallel$ [100], a small jump at $T_{\rm 0}$ in $C$($T$)/$T$ changes negligibly upon applying $B$ of 6 T in spite of an increase in $T_{\rm 0}$. 
In the longitudinal MR measurements for $B$ $\parallel$ [100], we found a sign change in the MR from negative to positive across $T_{\rm 0}$ and a $B$-induced anomaly at $B_x$ below $T_0$. 
The $M$($B$) curves for $B$ $\parallel$ [100] also exhibit small $B$-induced anomalies in both the HO and AFM states. 
These results suggest the presence of $B$-induced regions not only in the HO phase but also in the AFM phase in $B$--$T$ phase diagrams for $B$ $\parallel$ [100], which is in good agreement with the phase diagram constructed on the basis of the previous XRD measurements up to $B$ = 6 T. \cite{XRD_single} 
The presence of the $B$-induced regions may be a consequence of a change in the symmetry of the OP or domain alignment upon the application of $B$, which could be a clue to unraveling the mystery of the HO and the magnetic structure of the AFM state. 
Furthermore, a possibility that the $B$--$T$ phase diagram for $B$ $\parallel$ [001] also has the $B$-induced regions in both the ordered phases is suggested from the present $M$ measurements. 
However, the transverse MR for $B$ $\parallel$ [001] shows no anomalies and satisfies Kohler's rule above $T_{\rm N}$. 
Further verification is required to establish the $B$-induced regions for \mbox{$B$ $\parallel$ [001]}.

\vspace{2mm}

\begin{acknowledgment}
$\textbf{Acknowledgment}$ The authors are grateful to T. Matsumura, M. Manago, H. Kotegawa, H. Harima, and H. Kusunose for their helpful comments. 
The present research was supported by JSPS KAKENHI Grant Numbers, JP15H05882, JP15H05885(J-Physics), JP18H03683(A), and JP20K03825(C). 
This study was also partly supported by Hokkaido University, Global Facility Center (GFC), Advanced Physical Property Open Unit (APPOU), funded by MEXT under “Support Program for Implementation of New Equipment Sharing System” Grant Numbers JPMXS0420100316, JPMXS0420100317, JPMXS0420100318, and JPMXS0420100319. 
\end{acknowledgment}


\begin{thebibliography}{9}
\bibitem{Comment_Yatsushiro} M. Yastsushiro and S. Hayami theoretically indicated that an AFQ ordered state can be a ground state in CeCoSi by the mean-field calculation assuming a $\Gamma_7$ CEF GS and $\Gamma_6$ first excited state (see Ref. 8).
\bibitem{Comment_Nikitin} S. E. Nikitin $et$ $al$. also reported the $C$($T$) data for a polycrystalline CeCoSi, but did not comment on the anomaly at $T_{\rm 0}$ (see Ref. 10).
\end{thebibliography}
\end{document}